\documentclass{Odyssey2026}

 \interspeechcameraready

\title{Advancing Zero-Shot Open-Set Speech Deepfake Source Tracing}
\author[affiliation=1]{Manasi}{Chhibber}
\author[affiliation=1]{Jagabandhu}{Mishra}
\author[affiliation=1]{Tomi H.}{Kinnunen}

\affiliation{}{University of Eastern Finland}{Finland}

\email{manasi.chhibber@uef.fi, jagabandhu.mishra@uef.fi, tomi.kinnunen@uef.fi}

\usepackage{comment}
\usepackage{soul}
\usepackage{multirow} 
\begin{document}

\maketitle

\begin{abstract}
    
We propose a novel zero-shot source tracing framework inspired by speaker verification. We adapt SSL-AASIST for attack classification, enhancing embeddings with AAM loss and RegMixup, and ensure that training attacks are disjoint from those forming fingerprint–trial pairs. For backend scoring in attack verification, we explore both zero-shot approaches (cosine similarity and Siamese) and few-shot approaches (MLP and Siamese). Experiments on our recently introduced STOPA dataset with an open set setting show that few-shot learning provides advantages in the in-distribution (ID) scenario, while zero-shot approaches perform better in the out-of-distribution (OOD) scenario. In attack source verification with ID trials, few-shot Siamese and MLP achieve equal error rates (EER) of $17.72\%$ and $13.11\%$, compared to $29.91\%$ for zero-shot cosine scoring. Conversely, in OOD trials, zero-shot cosine scoring reaches $16.43\%$, outperforming few-shot Siamese $23.47\%$ and MLP $21.57\%$.

%We propose a novel zero-shot source tracing framework inspired by speaker verification. We adapt SSL-AASIST for attack classification, enhancing embeddings with AAM loss and RegMixup, and ensure that training attacks are disjoint from those forming fingerprint–trial pairs. For backend scoring, we explore zero-shot (cosine, Siamese) and few-shot (MLP, Siamese) approaches. Experiments on the STOPA dataset in an open-set setting show that few-shot methods excel in in-distribution trials(EER: Siamese $17.72\%$, MLP $13.11\%$), while zero-shot cosine scoring outperforms in out-of-distribution trials (EER $16.43\%$ vs. $23.47\%$ and $21.57\%$).
\end{abstract}

\section{Introduction}
\label{sec:intro}

\textit{“Trust, once lost, is not easily regained.”} Advances in neural speech synthesis and voice conversion now enable the creation of highly realistic spoofed speech~\cite{yi2023audio}. Such speech is often indistinguishable from bonafide human speech, both for listeners and for automatic systems~\cite{li2025survey}. The research community has responded with increasingly powerful spoofing detection models~\cite{li2025survey}. However, the challenge goes beyond simply detecting whether an utterance is spoofed. Forensics, accountability, and system security require attribution of a spoofing attack to its underlying generation method~\cite{klein2024source}. To address this need, the community organized a special session at Interspeech 2025, where the task of identifying the source of spoofing attacks was introduced as \emph{source tracing}~\cite{zhu2022source}.

\emph{Source tracing} is inherently an open-set multiclass classification task. This arises from the rapid advancement of generative methods, which continually introduce new spoofing attacks and thereby render the attack space unbounded~\cite{zhu2022source}. In the literature, the open-set source tracing task has been addressed broadly in two ways. The first approach focuses on detecting \emph{in-distribution} versus \emph{out-of-distribution} (OOD) attacks, followed by source identification of the in-distribution attacks. For example, \cite{falez2025audio} proposed an encoder–decoder architecture, while \cite{kulkarni25_interspeech} introduced a deep metric learning framework with multi-class N-pair loss and employed the Fréchet distance for joint evaluation of OOD attack detection and in-distribution attack source identification. Other contributions include a training-free method based on self-supervised (SSL) embeddings \cite{stan25_interspeech}, softmax energy-guided training for OOD attack detection \cite{klein25_interspeech}, continual learning strategies \cite{xiao25c_interspeech}, and variational information bottleneck methods \cite{doan25_interspeech}. A key challenge with these approaches is the lack of a unified metric for evaluating open-set source tracing performance.

\begin{figure}[t]
    \centering
    \includegraphics[width=\columnwidth] {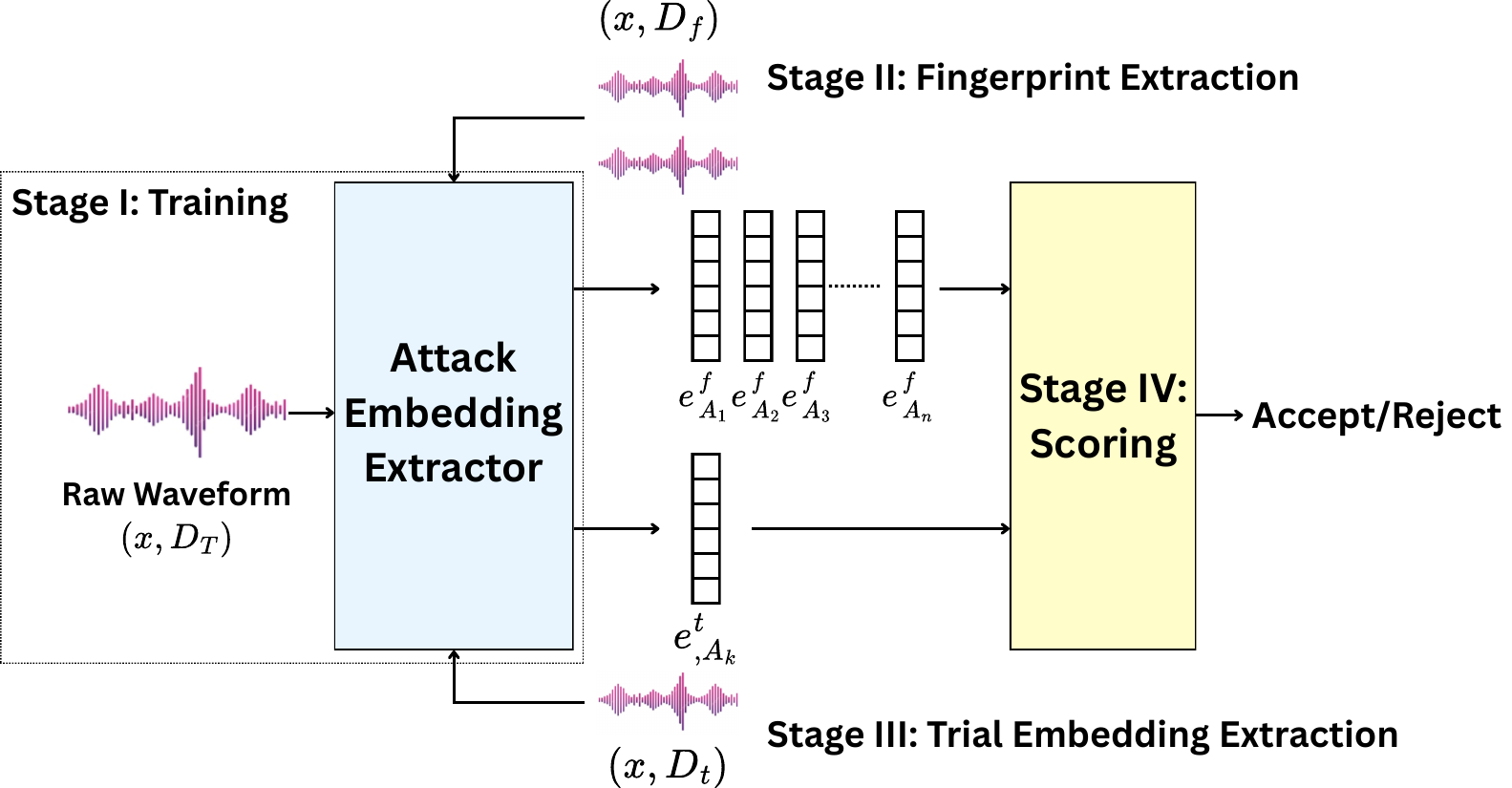}
    \caption{Overview of the proposed attribution framework, where a trial embedding derived from an attack utterance is compared against enrolled fingerprints to attribute the utterance to its underlying attack type through backend scoring.}
    \label{fig:intro_fig}
    \vspace{-0.4 cm}
\end{figure}

The second line of work addresses open-set source tracing by drawing inspiration from speaker verification~\cite{DODDINGTON2000225}, which inherently handles open-set problems. In this direction, \cite{koutsianos25_interspeech} applied metric learning with an SSL backbone, \cite{negroni25_interspeech} employed a ResNet18~\cite{he2016deep} based model, and \cite{falez2025audio} trained embedding extractors on seen attacks for source classification, later reusing the embeddings for source verification. Building on this idea, \cite{firc25_interspeech}, we introduced a systematic variation of spoofing audio for an open-set source tracing database (STOPA) and its associated evaluation protocol. The key design is to train attack embedding extractors on attacks that are completely \emph{disjoint} from those used during evaluation. Similar to speaker verification, where deep speaker embeddings serve as %fingerprint 
representations for enrolled and trial speakers, the STOPA evaluation protocol treats the enrollment phase as an \emph{attack fingerprinting} stage. During evaluation, trial utterances—including OOD attacks under the open-set protocol—are compared against these fingerprints to make attribution decisions. An illustration of this procedure is provided in Fig.~\ref{fig:intro_fig}. The pilot results reported in~\cite{firc25_interspeech} 
($39.1\%$ in-distribution and $35.3\%$ OOD EER) demonstrate the difficulty of the setting. 

\begin{figure*}[t]
    \centering
    \includegraphics[width=\textwidth]{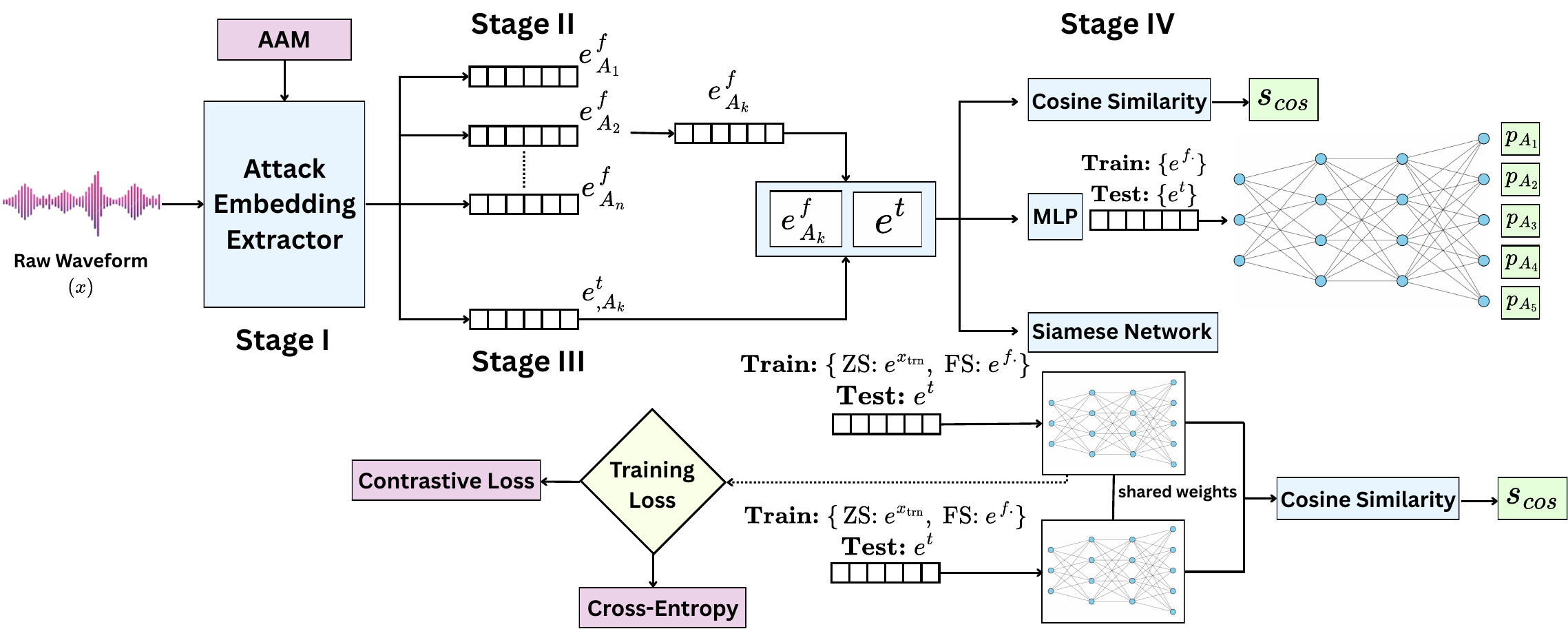}
    \caption{Verification-style source tracing framework. \textbf{Stage I} extracts embeddings from raw waveforms using a front-end model. In \textbf{Stage II}, fingerprints are formed by aggregating embeddings from multiple utterances of the same attack source. \textbf{Stage III} extracts embeddings for trial utterances. In \textbf{Stage IV}, backend scoring is applied: in the zero-shot (ZS) protocol, similarity is computed directly using cosine or a Siamese network trained on disjoint attacks; in the few-shot (FS) protocol, classifiers such as MLPs or Siamese networks are trained on fingerprint embeddings to refine decision boundaries. The framework naturally supports both closed-set and open-set source tracing.}
    \label{fig:full_arch}
\end{figure*}

In this work, we extend upon~\cite{firc25_interspeech} by introducing an improved training strategy for the embedding extractor and a more effective backend scoring. Specifically, we integrate a self-supervised front-end~\cite{wang2021investigating} to enhance generalization and improve robustness in capturing attack-specific characteristics. We further employ the additive angular margin (AAM) loss~\cite{deng2019arcface}, which has been shown to be well-suited for verification tasks due to its discriminative embedding properties. In addition, we incorporate regression mixup (RegMixup)~\cite{xie24_interspeech} as a regularization technique to improve model robustness and mitigate overfitting. Finally, we augment the training data with out-of-domain samples from ASVspoof2019~\cite{wang2021investigating}, thereby increasing attack diversity and enhancing the generalization capability of the learned attack fingerprint embeddings.

We also introduce a Siamese network as a backend scoring for source verification, supporting both zero-shot (trained on the embedding extractor training data) and few-shot (trained on fingerprint attack enrollment data) scenarios. In addition, we also introduce a few-shot MLP-based backend scoring approach. Finally, we conduct a comparative analysis of all introduced backend scoring methods against the standard zero-shot cosine scoring baseline.

\section{Open-set Attack Source Verification} 
\label{sec:format}
\subsection{Attack Source Verification}

\noindent \textbf{Source tracing} aims to identify or verify the source of a spoofing attack given an unknown utterance $x$. In an \emph{identification} setting, a system $\mathcal{F}_I$ predicts the source as
\begin{equation}
 \hat{k} = \arg\max_{k \in \mathcal{A}_{\text{ID}}} \mathcal{F}_I(x)_k,   
\end{equation}
where $\mathcal{A}_{\text{ID}}$ denotes the set of in-distribution (seen or known) attacks. In the \emph{closed-set} case, the source is always assumed to belong to $\mathcal{A}_{\text{ID}}$, whereas in the \emph{open-set} case, the system must also reject OOD attacks. One strategy, which avoids designing separate OOD classifiers, is to threshold the maximum score, i.e. $\max_{k} \mathcal{F}_I(x)_k < \tau$, where $\tau$ is a preset rejection threshold. 

In a \emph{verification} setting, the task is to accept or reject a claimed source $k$ by computing a similarity score $\psi(\cdot)$ between embeddings:  
\begin{equation}
\mathcal{F}_V(x, k) = 
\begin{cases}
\text{accept}, & \text{if } \psi(\vec{e}^{\,t}, \vec{e}^{\,f}_{\mathcal{A}_k}) \geq \tau \\
\text{reject}, & \text{otherwise},
\end{cases}
\end{equation}
where $\vec{e}^{\,t} = \zeta(x)$ is the trial embedding of utterance $x$, which may correspond to either an in-distribution or an out-of-distribution (OOD) attack, and $\zeta(\cdot)$ denotes the trained attack embedding extractor. The fingerprint embedding $\vec{e}^{\,f}_{\mathcal{A}_k}$ of the claimed source $\mathcal{A}_k$ is obtained as
\[
\vec{e}^{\,f}_{\mathcal{A}_k} = \phi\big(\zeta(x_{\mathcal{A}_k})\big),
\]  
where $\phi(\cdot)$ aggregates the embeddings of multiple utterances from attack $\mathcal{A}_k$ into a single fingerprint representation.

Overall, the framework (see Fig.~\ref{fig:full_arch}) consists of three stages: (1) training the attack embedding extractor $\zeta(\cdot)$, (2) constructing attack fingerprint embeddings $\vec{e}^{\,f}_{\mathcal{A}_k}$, and (3) backend scoring and verification. Let $\mathcal{D}=\{\mathcal{D}_T,\mathcal{D}_f,\mathcal{D}_t\}$ denote the dataset partitions, where $\mathcal{D}_T$ is used for training the attack embedding extractor, $\mathcal{D}_f$ for fingerprinting, and $\mathcal{D}_t$ for trials. The attack types $A^{T}_{\cdot}$ in $\mathcal{D}_T$ are disjoint from those in $\mathcal{D}_f$ and $\mathcal{D}_t$, with the relation $A^{f}_{\cdot} \subset A^{t}_{\cdot}$. 

\subsection{Attack Embedding Extractor Training}

The embedding extractor $\zeta(\cdot)$ is trained on utterances and their corresponding spoofing attack labels 
$\{(x_i, \mathcal{A}_i)\}_{i=1}^{N_T}$ using the training set $\mathcal{D}_T$. To enhance between-attack discrimination of the embeddings and improve the generalization of the attack embedding extractor, we adopt a combined training objective consisting of AAM softmax loss and RegMixup. The AAM softmax loss is defined as,
\begin{equation}
\mathcal{L}_{\text{AAM}} = - \frac{1}{N_T} \sum_{i=1}^{N_T} 
\log \frac{e^{s \cdot (\cos(\theta_{y_i} + m))}}{ e^{s \cdot (\cos(\theta_{y_i} + m))} + \sum_{j \neq y_i} e^{s \cdot \cos \theta_j} },
\end{equation}
\noindent where $y_i$ is the ground-truth attack label of $x_i$, $\theta_j$ denotes the angle between the embedding $\mathbf{e}_i$ and the weight vector corresponding to attack class $j$, $m$ is the additive angular margin, and $s$ is a scaling factor. This objective encourages embeddings of the same attack to cluster tightly while increasing the angular separation between embeddings of different attacks, thereby improving generalization to both in-distribution and out-of-distribution (OOD) attacks.

%To further improve robustness and enforce smoother decision boundaries in the embedding space, we incorporate regression mixup (RegMixup)~\cite{}.

Given two training samples $(x_i, y_i)$ and $(x_j, y_j)$ for incorporating RegMixup~\cite{xie24_interspeech}, we construct a virtual training example
\begin{equation}
\tilde{x} = \lambda x_i + (1-\lambda)x_j, \qquad
\tilde{y} = \lambda y_i + (1-\lambda)y_j,
\end{equation}
\noindent where $\lambda \sim \text{Beta}(\alpha,\alpha)$ controls the interpolation strength. The mixed sample $\tilde{x}$ is passed through the embedding extractor, and the resulting representation is optimized using the same AAM objective with the interpolated target $\tilde{y}$. The overall training objective is thus given by
\begin{equation}
\mathcal{L} = \mathcal{L}_{\text{AAM}}(x, y) + \mathcal{L}_{\text{AAM}}(\tilde{x}, \tilde{y}),
\end{equation}
which promotes more discriminative and smoothly varying representations across attack classes, leading to improved robustness and generalization to OOD attacks.

% The embedding extractor $\zeta(\cdot)$ is trained on utterances and their corresponding spoofing attack labels 
% $\{(x_i, \mathcal{A}_i)\}_{i=1}^{N_T}$ using the training set $\mathcal{D}_T$. To enhance 
% between-attack discrimination of the embeddings, we adopt the additive angular margin (AAM) softmax loss: 
% \begin{equation}
% \mathcal{L}_{\text{AAM}} = - \frac{1}{N_T} \sum_{i=1}^{N_T} 
% \log \frac{e^{s \cdot (\cos(\theta_{y_i} + m))}}{ e^{s \cdot (\cos(\theta_{y_i} + m))} + \sum_{j \neq y_i} e^{s \cdot \cos \theta_j} },
% \end{equation}
% \noindent where $y_i$ is the ground-truth attack label of $x_i$, $\theta_j$ is the angle between $\vec{e}_i$ and the weight vector of attack $j$, $m$ is the additive angular margin, and $s$ is a scaling factor. This objective encourages embeddings of the same attack to cluster tightly while increasing the angular separation between embeddings of different attacks, improving generalization for both in-distribution and OOD attacks.

\subsection{Attack Fingerprint Extraction}

For each attack source $\mathcal{A}_k$, we first select a total of $r$ utterances from the fingerprint dataset $\mathcal{D}_f$. These reference utterances are used to compute a representative fingerprint embedding for the attack. For each selected utterance $x_i$, the embedding is extracted using the trained extractor $\zeta(\cdot)$.
The fingerprint embedding $\vec{e}^{\,f}_{\mathcal{A}_k}$ is then obtained by aggregating these embeddings using the mean:
\begin{equation}
\vec{e}^{\,f}_{\mathcal{A}_k} = \phi\Big(\{\vec{e}_i\}_{i=1}^r\Big) = \frac{1}{r} \sum_{i=1}^{r} \vec{e}_i,
\end{equation}
\noindent 
which ensures that the fingerprint embedding captures a representative attack signature while reducing variability across individual utterances. By varying $r$, we can analyze the impact of the number of fingerprint utterances on verification performance in both few-shot and zero-shot scenarios.

\subsection{Zero-shot and Few-shot Backend Scoring}

For verification, trial utterances are drawn from the trial set $\mathcal{D}_t$ and compared against fingerprint embeddings computed from $\mathcal{D}_f$. We consider four backend scoring strategies: (1)   \textbf{Zero-shot cosine:} Simple cosine similarity between the trial embedding $\vec{e}^{\,t}$ and the fingerprint embedding $\vec{e}^{\,f}_{\mathcal{A}_k}$, (2) \textbf{Zero-shot Siamese:} A Siamese network trained on $\mathcal{D}_T$, with attack types disjoint from the fingerprint set, following the zero-shot protocol, (3) \textbf{Few-shot Siamese:} A Siamese network trained on fingerprint pairs from $\mathcal{D}_f$, and (4) \textbf{Few-shot MLP:} An MLP classifier trained on attack embeddings from $\mathcal{D}_f$, targeting in-distribution attacks.  

In the \textbf{zero-shot} cosine setting, the similarity score is defined as
% \begin{equation}
% \psi(\vec{e}^{\,t}, \vec{e}^{\,f}_{\mathcal{A}_k}) = 
% \frac{\vec{e}^{\,t} \cdot \vec{e}^{\,f}_{\mathcal{A}_k}}{\|\vec{e}^{\,t}\| \, \|\vec{e}^{\,f}_{\mathcal{A}_k}\|}.
% \label{eq:cosine_score}
% \end{equation}

\begin{equation}
\psi(\vec{e}^{\,t}, \vec{e}^{\,f}_{\mathcal{A}_k}) =
\frac{\vec{e}^{\,t} \cdot \vec{e}^{\,f}_{\mathcal{A}_k}}
{\lVert \vec{e}^{\,t} \rVert \, \lVert \vec{e}^{\,f}_{\mathcal{A}_k} \rVert}.
\label{eq:cosine_score}
\end{equation}

For the Siamese networks, $N$ trial–fingerprint pairs with equal positive and negatives are generated from the respective training partition. The network is trained using contrastive learning~\cite{he2020momentum}. After training, verification scores are computed using the cosine similarity measure between embeddings, as defined in Eq.~\ref{eq:cosine_score}. 

For the \textbf{few-shot MLP}, cross-entropy is used to train on attack embeddings from $\mathcal{D}_f$, and during evaluation, the output probability corresponding to the claimed attack is taken as the verification score.

\section{Experimental Setup}
\label{sec:majhead}

% \begin{figure*}[t]
% % \vspace{-0.5cm}
%     \centering
%     \includegraphics[width=\textwidth]{od2026_latex_template/images/SSL-ResNet+AAM.pdf} 
%     \caption{SSL-ResNet + AAM}
% % \vspace{-0.6 cm}
%     \label{fig:eer}
% \end{figure*}

% \begin{figure*}[t]
% % \vspace{-0.5cm}
%     \centering
%     \includegraphics[width=\textwidth]{od2026_latex_template/images/SSL-AASIST+AAM+mixup.pdf} 
%     \caption{SSL-AASIST + AAM + mixup}
% % \vspace{-0.6 cm}
%     \label{fig:eer}
% \end{figure*}

\subsection{Database}

We conduct experiments primarily on the recent, publicly available STOPA~\cite{firc25_interspeech} dataset. It contains $699$k spoofed utterances from $13$ attack systems, formed by combining $8$ acoustic models (AMs) and $6$ vocoder models (VMs). Each utterance is labeled with its attack id as well as AM and VM ids, enabling multi-level source tracings. Following the open-set protocol, three attacks (corresponding to 20 speakers) 
are used to train the embedding extractor ($\mathcal{D}_T$) and 
five attacks (from 10 speakers)  
are used for fingerprinting ($\mathcal{D}_f$). The remaining five attacks (48 speakers), along with the five attacks used in fingerprinting, are used in the evaluation set ($\mathcal{D}_t$). The speakers are disjoint across all partitions. In the fingerprinting stage, $r \in \{1, 10, 100, 400\}$ utterances are selected for common sentences (co) having identical linguistic content, and $r \in \{1, 10, 100, 400, 1000, \text{all}\}$ for non-common sentences (nc).  

To diversify embedding extractor training, we include an additional 121k utterances from six attacks (A01–A06) in the ASVspoof 2019 LA~\cite{wang2020asvspoof} corpus into $\mathcal{D}_T$. This results in a total of nine training attacks across the two datasets. All utterances are preprocessed with silence trimming to remove non-speech regions.

\subsection{Experimental Setup}
We employ SSL-AASIST~\cite{tak2022automatic} due to its strong adaptability to both spoofing detection and source tracing tasks, along with SSL-ResNet34~\cite{dao24_asvspoof}, which has demonstrated promising performance in ASVspoof5. The performance of these models is compared against a baseline AASIST model trained solely on STOPA~\cite{firc25_interspeech}. For training, the partition $\mathcal{D}_T$ is randomly divided into training and validation sets in an $80:20$ ratio.

\begin{table*}
\centering
\caption{EERs (\%) for different embedding extractors. 
AASIST* denotes a model trained solely on STOPA, while the rest are trained on STOPA combined with ASVspoof 2019 training set. The latter achieves the best generalization to OOD attacks. AM: acoustic model, VM: vocoder model, ID: in-distribution.}
\label{tab:table_1}
\begin{tabular}{l ccc ccc}
\toprule
\multirow{2}{*}{\textbf{Embedding Extractor (\#Params)}} & 
\multicolumn{3}{c}{\textbf{ID}} & 
\multicolumn{3}{c}{\textbf{OOD}} \\
\cmidrule(lr){2-4} \cmidrule(lr){5-7}
 & \textbf{Attack} & \textbf{AM} & \textbf{VM} 
 & \textbf{Attack} & \textbf{AM} & \textbf{VM} \\
\midrule
AASIST* (301K) & 47.05 & 47.05 & 45.25 & 47.75 & 49.37 & 48.68 \\
SSL-AASIST (318M) & 37.69 & 37.69 & 37.56 & 42.28 & 44.37 & 40.21 \\
SSL-ResNet34 + AAM (43M) & 39.76 & 39.76 & 38.65 & 26.08 & 41.79 & 40.40 \\
SSL-AASIST + AAM (318M) & 38.39 & 38.39 & 35.88 & 33.13 & 36.31 & 34.07 \\
SSL-AASIST + AAM + RegMixup (318M) & \textbf{31.29} & \textbf{31.29} & \textbf{28.80} & \textbf{21.43} & \textbf{36.85} & \textbf{27.15} \\
% SSL-AASIST + AAM (318M) & \textbf{38.39} & \textbf{38.39} & \textbf{35.88} & \textbf{33.13} & \textbf{36.31} & \textbf{34.07} \\
\bottomrule
\end{tabular}%
\end{table*}
\begin{figure*}[h!]
    \vspace{-0.1cm} 
    \centering
    \includegraphics[height= 140pt,width=470pt]{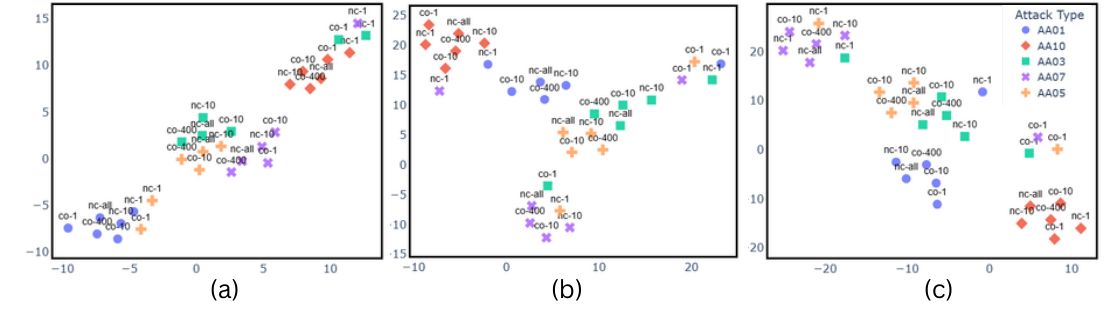}
    \caption{Scatter plot of averaged fingerprint embeddings using t-SNE: (a) ResNet34+AAM, (b) SSL-AASIST+AAM, (c) SSL-AASIST+AAM+RegMixup, shown for different enrollment sizes. Larger enrollments yield tighter clusters and improved separability across attack types, highlighting the importance of enrollment scale for attribution accuracy. Notably, RegMixup further smooths and regularizes the embedding space, producing more compact and well-separated clusters.}
    \label{fig:scatter}
\end{figure*}

%For training, the partition $\mathcal{D}_T$ is randomly split $80:20$ for training and validation.

We train both SSL-AASIST and SSL-ResNet34 using the AAM-softmax objective with a scaling factor of $30$ and an additive margin of $0.5$. We train both models for $100$ epochs using the Adam optimizer with a learning rate of $0.0001$, and set the RegMix interpolation parameter $\alpha$ to $0.2$. We design the Siamese network with three linear layers of dimensions $128 \rightarrow 64 \rightarrow 32$, and train it for $100$ epochs on $50,000$ positive and negative pairs using the Adam optimizer with a learning rate of $0.001$. We configure the MLP backend with one hidden layer of $128$ units and train it for $100$ epochs using the Adam optimizer with a learning rate of $0.001$. We tune model size and hyperparameters on the validation set based on validation loss, and select the model with the lowest validation loss for subsequent processing. We make our implementation publicly available on GitHub for reproducibility.\footnote{https://github.com/Manasi2001/Zero-Shot-Open-Set-Speech-Deepfake-Source-Tracing}

We evaluate all experiments using the Equal Error Rate (EER) under two conditions: in-distribution  and OOD. The in-distribution EER is computed from target and non-target trials within the in-distribution set. The OOD EER, on the other hand, is computed by contrasting in-distribution target trials with OOD non-target trials.

% SSL-AASIST is trained with an AAM-softmax objective, using a scaling factor of $30$ and an additive margin of $0.5$. The model is trained for $100$ epochs using the Adam optimizer with a learning rate of $0.0001$. The Siamese network consists of $3$ linear layers with hidden dimensions $128 \rightarrow 64 \rightarrow 32$, and is trained for $100$ epochs on $50,000$ positive/negative pairs from the corresponding training set using the Adam optimizer with a learning rate of $0.001$. The MLP backend has $1$ hidden layer with $128$ units and is trained for $100$ epochs using the Adam optimizer with a learning rate of $0.001$. Model size and hyperparameters are optimized on the validation set with respect to the validation loss. The model achieving the lowest validation loss is selected for subsequent processing. The implementation of our work is available on GitHub for reference.\footnote{https://github.com/Manasi2001/Zero-Shot-Open-Set-Speech-Deepfake-Source-Tracing}

\section{Results and Discussion}

\subsection{Zero-shot Attack Source Verification}
We report the attack source, acoustic model, and vocoder model verification EERs using zero-shot cosine scoring in Table~\ref{tab:table_1}. 

We first compare the baseline AASIST model, trained on only three attacks, with SSL-AASIST trained using cross-entropy and AAM-softmax. We observe that SSL-AASIST consistently outperforms the baseline across all scenarios, demonstrating the effectiveness of incorporating a self-supervised front end. Furthermore, we show that augmenting the training data with additional attacks from the ASVspoof corpus further improves performance. While AAM-softmax yields performance comparable to cross-entropy for in-distribution acoustic and vocoder model verification, it provides comparatively better results for OOD attacks.

We compare SSL-AASIST with SSL-ResNet34 under the AAM-softmax setting. In the in-distribution scenario, SSL-AASIST achieves superior performance, whereas in the OOD scenario, SSL-ResNet34 performs better for attack source verification. We attribute this improvement to the relatively smaller model size of SSL-ResNet34 (43M parameters compared to 318M in SSL-AASIST), which likely enhances generalization by reducing overfitting. However, SSL-AASIST consistently outperforms SSL-ResNet34 in acoustic and vocoder model verification tasks. This suggests that although SSL-ResNet34 embeddings generalize well for attack source verification, they fail to adequately capture acoustic and vocoder-specific characteristics. Further, we visualize the embedding spaces of the fingerprint representations through scatter plots in Fig.~\ref{fig:scatter} (a) and (b) using t-distributed stochastic neighbor embedding (t-SNE) for SSL-ResNet34+AAM and SSL-AASIST+AAM. Apart from the one-utterance ($r=1$) case, the fingerprints form compact clusters, with SSL-AASIST producing noticeably better clusters compared to SSL-ResNet34.

Finally, we incorporate RegMixup into SSL-AASIST training and observe performance improvements across attack, acoustic model, and vocoder model verification in both in-distribution and OOD scenarios, although acoustic model verification remains largely comparable. Notably, the improvements are more pronounced in the OOD setting, demonstrating that RegMixup enhances the generalization capability of the learned attack embeddings. We also visualize the fingerprint embedding space in Fig.~\ref{fig:scatter} (b) and (c). With RegMixup, attack fingerprints form tighter and more distinct clusters compared to training without it, resulting in a smoother and better-regularized embedding space. Notably, these fingerprints correspond to attacks that were not used to train the attack embedding extractor, highlighting the embedding extractor’s generalizability.

\begin{table*}
\centering
\caption{EERs(\%) under zero-shot and few-shot backend strategies with nc-all condition (as the MLP and Siamese in few-shot have seen all utterances) using SSL-AASIST+AAM+RegMixup. AM: acoustic model, VM: vocoder model, ID: in-distribution.}
\label{tab:table_2}
\begin{tabular}{llcccccc}
\toprule
\multirow{2}{*}{\shortstack{\textbf{Learning} \\ \textbf{Strategy}}} & 
\multirow{2}{*}{\textbf{Backend}} & 
\multicolumn{3}{c}{\textbf{ID}} & 
\multicolumn{3}{c}{\textbf{OOD}} \\
\cmidrule(lr){3-5} \cmidrule(lr){6-8}
 &  & \textbf{Attack} & \textbf{AM} & \textbf{VM} 
 & \textbf{Attack} & \textbf{AM} & \textbf{VM} \\
\midrule
Zero-shot & Cosine Similarity & 29.91 & 29.91 & 28.37 & \textbf{16.43} & 33.14 & \textbf{21.00}  \\
Zero-shot  & Siamese Network   & 42.21 & 42.21 & 54.20 & 42.74 & 41.93 & 46.55 \\ \hline
Few-shot  & MLP  & \textbf{13.11} & \textbf{13.11} & \textbf{13.55} & 21.57 & \textbf{29.83} & 25.21 \\
%Few-shot & Siamese Network (CE) & 16.38 & 16.38 & 15.89 & 24.03 & 30.39 & 26.61 \\
Few-shot & Siamese Network  & 17.72 & 17.72 & 14.95 & 23.47 & 31.90 & 24.94 \\
\bottomrule
\end{tabular}%

\end{table*}
% \begin{figure}[h!]
%     \centering
%     \includegraphics[height= 150pt,width=240pt]{od2026_latex_template/images/SSL-ResNet+AAM.png}
%     \caption{SSL-ResNet + AAM}
%     \label{fig:scatter}
% \end{figure}

% \begin{figure}[h!]
%     \centering
%     \includegraphics[height= 150pt,width=240pt]{od2026_latex_template/images/SSL-AASIST+AAM+mixup.png}
%     \caption{SSL-AASIST + AAM + mixup}
%     \label{fig:scatter}
% \end{figure}

% \begin{figure}[h!]
%     \centering
%     \includegraphics[height= 150pt,width=240pt]{od2026_latex_template/images/SSL-ResNet+AAM+mixup.png}
%     \caption{SSL-ResNet + AAM + mixup}
%     \label{fig:scatter}
% \end{figure}
\begin{figure*}[t]
    \centering
    \includegraphics[height= 190pt,width=480pt]{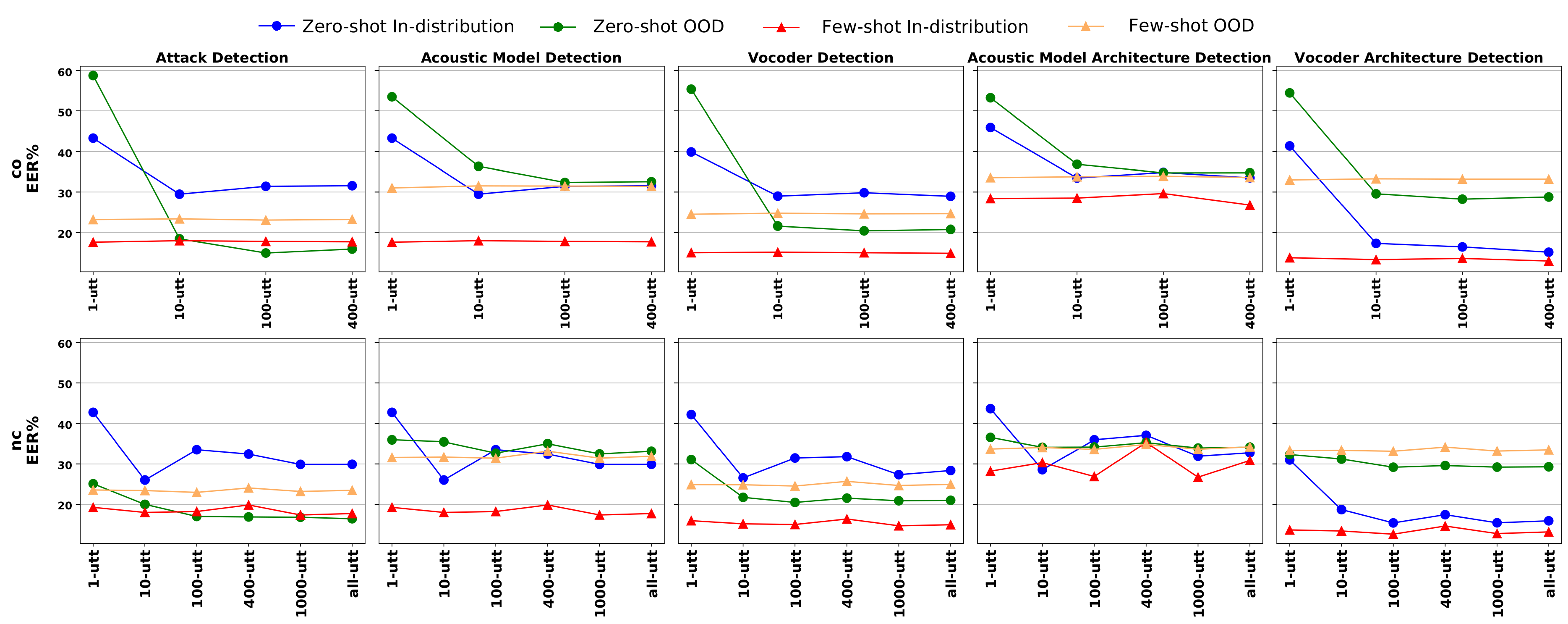} 
    \caption{SSL-AASIST + AAM + RegMixup: EER trends across attack-, acoustic model (AM)-, vocoder model (VM)-, and architecture-level verification for zero-shot and few-shot backends under in-distribution (ID) and out-of-distribution (OOD) scenarios. The top row shows common utterances (co) with identical linguistic content, while the bottom row shows non-common utterances (nc). Zero-shot results use cosine similarity scoring, and few-shot results use a Siamese network trained with contrastive loss.
}
    \label{fig:eer}
\end{figure*}

%The scatter plot in Fig.~\ref{fig:scatter} visualizes the embedding space of fingerprint representations generated by SSL-AASIST trained with the AAM-softmax loss. This plot highlights the discrimination ability of our attack embedding extractor. As observed, all attacks form well-separated clusters in the embedding space, with the exception of a single utterance case.

\subsection{Analysis of Backend Learning Strategies}

The embeddings generated using SSL-AASIST with AAM and RegMixup are used to train the backend scoring methods. The performance of these backends is reported in Table~\ref{tab:table_2}. In the zero-shot setting, cosine scoring outperforms the Siamese network across attack, acoustic model, and vocoder model verification. This may be because the Siamese network struggles to discriminate effectively and may require more trial pairs to improve performance. 

Comparing zero-shot and few-shot Siamese settings, the few-shot setup clearly improves performance across all verification tasks, with attack verification EER improving from $42.21\%$ to $17.72\%$. Using MLP in the few-shot setting further reduces attack source verification EER to $13.11\%$, demonstrating its effectiveness. However, for OOD attacks, zero-shot cosine scoring still outperforms the other backend methods, except in acoustic model source verification under OOD conditions, where the MLP achieves better performance. We also observe that attack source verification generally performs better than both acoustic and vocoder model source verification. Among these, acoustic model source verification is particularly challenging, performing worse than vocoder model source verification.

\subsection{Performance Analysis with Enrollment Size and Attack Source Granularity}

Fig.~\ref{fig:eer} presents EER comparisons for zero-shot cosine and few-shot Siamese backends under both \textit{co} and \textit{nc} conditions. In the in-distribution setting, few-shot verification consistently outperforms the other methods, whereas in the OOD setting, the trend reverses. Increasing the number of utterances used for fingerprinting consistently improves verification performance. We observe no notable differences between the \textit{co} and \textit{nc} conditions, suggesting that linguistic constraints play a minimal role.

At different levels of attack verification, vocoder models exhibit lower EERs than acoustic models. Moreover, both acoustic and vocoder model verification achieve lower EERs than their corresponding architecture-level verification (acoustic model–architecture and vocoder model–architecture), indicating that tracing across different architectures within the same model family is more challenging than tracing at the model level.

\section{Conclusion}

In this work, we addressed a realistic and challenging open-set, zero-shot source tracing scenario. We enhanced SSL-AASIST embeddings using AAM loss and RegMixup, incorporating out-of-distribution data to improve variability and robustness. In zero-shot tracing, cosine similarity generalized best to OOD attacks, while few-shot backends (MLP, Siamese) provided improvements in in-distribution scenarios. The Siamese backend was effective in few-shot settings but less robust under zero-shot conditions.

We also observed that increasing the number of utterances used for fingerprinting consistently improved verification performance. At the model level, attack source verification generally outperformed acoustic and vocoder model verification, with acoustic models being the most challenging to distinguish. Tracing attacks across shared architectures proved particularly difficult, emphasizing the limits of current embedding approaches. Visualization of fingerprint embeddings further confirmed that RegMixup produces more compact and well-separated clusters, demonstrating improved generalization of the learned embeddings. Overall, this work demonstrates the feasibility of zero-shot source tracing but also highlights significant performance gaps, motivating future efforts toward more robust and scalable methods.

\section{Acknowledgements}

The work has been partially supported by the Academy of Finland (Decision No. 349605, project "SPEECHFAKES"). The authors wish to acknowledge CSC – IT Center for Science, Finland, for computational resources.

% \ifinterspeechfinal
%      The Odyssey 2026 organisers
% \else
%      The authors
% \fi
% would like to thank ISCA and the organising committees of past Interspeech conferences for kindly providing the previous version of this template.

\bibliographystyle{IEEEtran}
\bibliography{main}

\end{document}